\documentclass[fleqn,usenatbib]{mnras}

\usepackage{newtxtext,newtxmath}
\usepackage[normalem]{ulem}

\usepackage[T1]{fontenc}
\usepackage[dvipsnames]{xcolor}
\usepackage{float}

\DeclareRobustCommand{\VAN}[3]{#2}
\let\VANthebibliography\thebibliography
\def\thebibliography{\DeclareRobustCommand{\VAN}[3]{##3}\VANthebibliography}

\usepackage{graphicx}	% Including figure files
\usepackage{amsmath}	% Advanced maths commands
\usepackage{orcidlink}

\title[IGM Phase Evolution]{Emergence of the Temperature-Density Relation in the Low-Density Intergalactic Medium}

\author[A. Wells et al.]{
Alexandra Wells \orcidlink{0000-0001-5080-4582},$^{1,2}$\thanks{E-mail: aw771620@ohio.edu}
David Robinson \orcidlink{0000-0002-3751-6145},$^{2,3}$
Camille Avestruz \orcidlink{0000-0001-8868-0810},$^{2,3}$
Nickolay Y. Gnedin \orcidlink{0000-0001-5925-4580}$^{4,5,6}$
\\
$^{1}$Physics and Astronomy Department; Ohio University, Athens, OH 45701, USA\\
$^{2}$Department of Physics; University of Michigan, Ann Arbor, MI 48109, USA\\
$^{3}$Leinweber Center for Theoretical Physics; University of Michigan, Ann Arbor, MI 48109, USA\\
$^{4}$Theoretical Physics Division; 
Fermi National Accelerator Laboratory;
Batavia, IL 60510, USA\\
$^{5}$Kavli Institute for Cosmological Physics;
The University of Chicago;
Chicago, IL 60637, USA\\
$^{6}$Department of Astronomy \& Astrophysics; 
The University of Chicago; 
Chicago, IL 60637, USA
}

\date{Accepted XXX. Received YYY; in original form ZZZ}

\pubyear{2023}

\begin{document}
\label{firstpage}
\pagerange{\pageref{firstpage}--\pageref{lastpage}}
\maketitle

\begin{abstract}
We examine the evolution of the phase diagram of the low-density intergalactic medium during the Epoch of Reionization in simulation boxes with varying reionization histories from the Cosmic Reionization on Computers project. The PDF of gas temperature at fixed density exhibits two clear modes: a warm and cold temperature mode, corresponding to the gas inside and outside of ionized bubbles. We find that the transition between the two modes is ``universal'' in the sense that its timing is accurately parameterized by the value of the volume-weighted neutral fraction for any reionization history. This ``universality'' is more complex than just a reflection of the fact that ionized gas is warm and neutral gas is cold: it holds for the transition at a fixed value of gas density, and gas at different densities transitions from the cold to the warm mode at different values of the neutral fraction, reflecting a non-trivial relationship between the ionization history and the evolving gas density PDF.
Furthermore, the ``emergence'' of the tight temperature-density relation in the warm mode is also approximately ``universally'' controlled by the volume-weighted neutral fraction for any reionization history. In particular, the ``emergence'' of the temperature-density relation (as quantified by the rapid decrease in its width) occurs when the neutral fraction is $10^{-4}\lesssim X_\mathrm{HI} \lesssim10^{-3}$ for any reionization history. Our results indicate that the neutral fraction is a primary quantity controlling the various properties of the temperature-density relation, regardless of reionization history.

\end{abstract}


%%%%%%%%%%%%%%%%%%%%%%%%%%%%%%%%%%%%%%%%%%%%%%%%%%

%%%%%%%%%%%%%%%%% BODY OF PAPER %%%%%%%%%%%%%%%%%%

\section{Introduction}

A second after the Big Bang, the universe was filled with a plasma of free protons and electrons. As the universe expanded over time, it also cooled enough that these protons and electrons combined to form neutral hydrogen atoms, decoupling from the primordial Black Body radiation known today as the cosmic microwave background (CMB) radiation \citep{p&w1965}. Immediately after this epoch, there were no other sources of light as large-scale structures had not yet been formed. This waiting period, often denoted the ``Dark Ages”, lasted until the first stars began to form and emit light some 100-200 million years after the Big Bang.

This emitted UV radiation was then able to ionize nearby regions, or ‘bubbles’, of hydrogen gas in what is known as the Epoch of Reionization (for a recent review, see \citet{wise19}). This process occurred approximately from $z=15$ to $z=6$. These bubbles continued to grow as more neutral hydrogen became reionized and the bubbles eventually began to overlap, until nearly the entire intergalactic medium (IGM) was ionized. 

It is known that the IGM at $z<6$ consists mostly of ionized hydrogen because of the observed absorption spectra of distant quasars. As (UV) photons from a quasar pass through a cloud of neutral hydrogen gas, an absorption line corresponding to the Lyman-$\alpha$ transition appears. From passing through multiple gas clouds at different redshifts, a series of absorption lines results called the Lyman-$\alpha$ forest. The fact that this absorption is not complete and that flux from these high redshift objects is transmitted is proof that the IGM hydrogen must be highly ionized, as fully neutral (or even neutral at the $\sim 10^{-3}$ level) IGM would absorb all Lyman-$\alpha$ photons \citep{g&p1965}.  

Simulations of the IGM show that the mean temperature of the IGM is approximately related to its density by a power-law relation, $T(\delta) = T_0(1+\delta)^{\gamma-1}$, where $T_0$ is the temperature at the mean density and $\gamma$ is the power-law slope \citep{Hui&Gnedin1997, Schaye1999, Ricotti2000, McQuinn2016}.

Measurements of the Ly$\alpha$ forest can thus be used to constrain the parameters of this temperature-density relation (TDR), $T_0$ and $\gamma$ \citep[e.g.][]{Boera2014, Boera2016, Nasir2016, Rorai2018, Walther2019, Muller2021}. Both $T_0$ and $\gamma$ are theoretically expected \citep[e.g.][]{Hui&Gnedin1997, Ricotti2000, Puchwein2012, Nasir2016, McQuinn2016, Puchwein2023} and have been observationally measured to evolve  \citep[e.g.][]{Chang2012, Boera2014, Hiss2018, Walther2019}.

While a number of studies have found $\gamma > 1$ between $1 \lesssim z \lesssim 5$ (after reionization) so that temperature increases with increasing density \citep{Boera2014, Rorai2018, Walther2019, Muller2021}, there has also been observational evidence for an ``inverted'' TDR with $\gamma < 1$ at low density and low redshifts \citep[e.g.][]{Bolton2008, Puchwein2012, Chang2012}, or a broken power law where the TDR is inverted for densities lower than about twice the mean density \citep{Rorai2017}. Such an inverted TDR can only be produced by non-radiative heating sources, for example, ultra-high energy cosmic rays from blazars driving plasma instabilities in the IGM \citep[e.g.][]{Chang2012, Puchwein2012}. 

Hence, further theoretical understanding of how the TDR of the low-density IGM emerges during the Epoch of Reionization is needed to better interpret measurements of the thermal state of the IGM post-reionization. Recently, simulations that track the post-reionization thermal state of the IGM have been performed, which both do \citep{Puchwein2023} and do not \citep{Villasenor2021} explicitly model the reionization process, but the actual process of ``emergence'' of the TDR during reionization is not yet well understood.

In this paper, we use existing Epoch of Reionization simulations to study the timing of the thermal evolution of the low-density IGM, including when the temperature-density relation emerges. First, we introduce the simulation project used in this study and show how the phase diagrams demonstrate the IGM evolution. We then describe the separation of two thermal modes in this evolution, analyze the timing of the transition between these modes, and look at how the evolution of the neutral fraction is related to this timing. We examine the temperature-density relation that emerges post-reionization and see how its slope and FWHM change with time and across reionization histories.

\section{Methodology}

\subsection{CROC Simulations}
\label{methods:CROC}

\begin{figure}
    \includegraphics[width=\columnwidth]{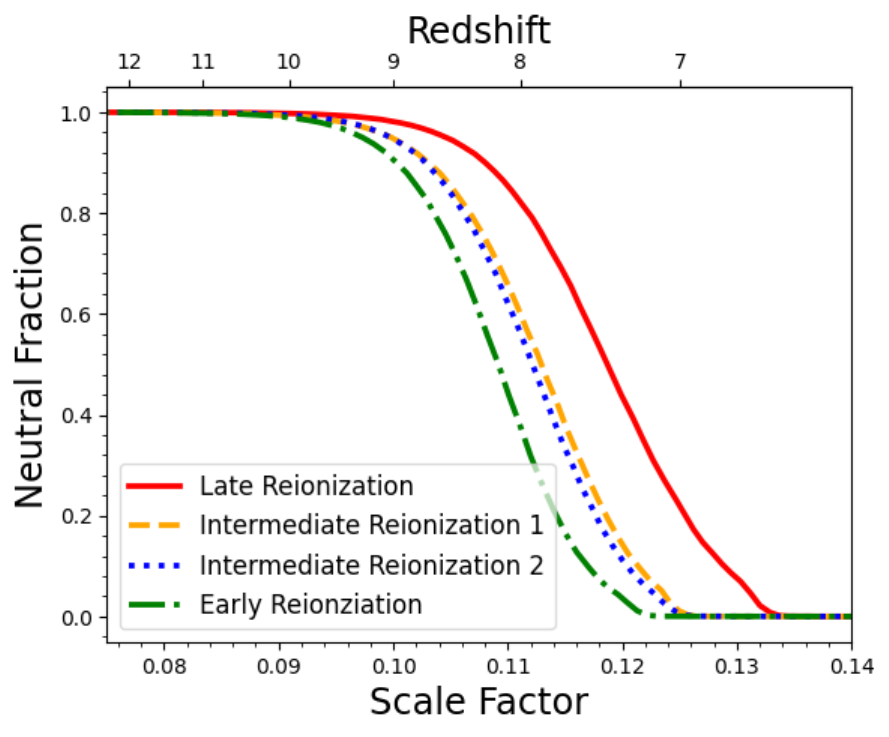}
    \caption{The volume-weighted neutral fraction of the IGM as a function of scale factor at different simulation boxes. The four boxes we use sample the full range of reionization histories in CROC simulations: the latest reionization time (solid line), two boxes with similar intermediate reionization times (dash and dot lines), and the earliest reionization time (dash-dotted line).}
    \label{fig:rei-history}
\end{figure}

In this study, we use numerical simulations from the ``Cosmic Reionization on Computers'' (CROC) project (to learn more about the CROC project, see \citet{CROC1} and \citet{CROC2}). These simulations model self-consistently all physical processes generally considered to be relevant to cosmic reionization, such as radiative transfer, star formation and feedback, and ionizing radiation from stars and (approximately) from quasars. We use four 40$h^{-1}$ comoving Mpc boxes which sample the full range of reionization histories modeled by the CROC project. Fig.~\ref{fig:rei-history} shows the reionization histories of the different simulation boxes. These four models differ only by the value of the rms density fluctuation at the box scale (the so-called "DC mode") but otherwise include the same physical modeling and are simulated with the same numerical parameters.

\subsection{IGM Evolution}

\begin{figure*}

	\includegraphics[width=6in]{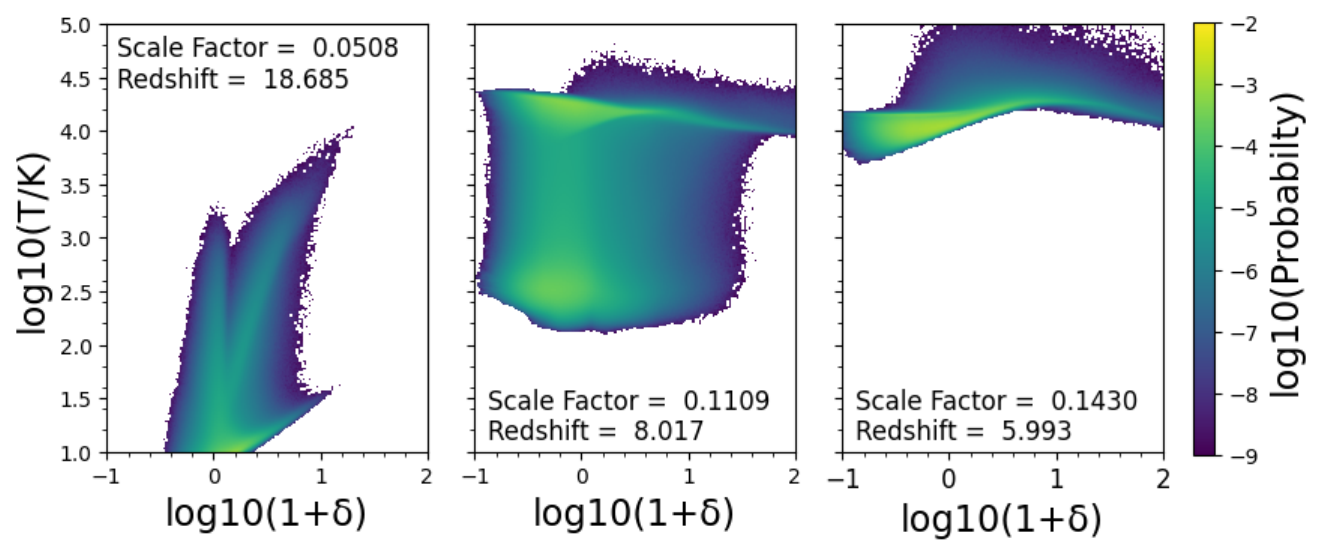}
    \caption{The temperature-density phase diagram at three different scale factors, which shows the temperature evolution of the IGM over the Epoch of Reionization for a simulation with intermediate reionization history 1. The first panel shows the IGM at an early scale factor of $a= 0.05$, where the neutral fraction is $0.9998$. Most of the IGM is at a temperature below $30$ K. The middle panel shows the IGM at a scale factor $a = 0.111$, where the 'cold mode' and 'warm mode' fractions of the IGM are approximately equal, and the neutral fraction is $0.6$. The third panel shows the IGM at a higher scale factor of $a = 0.143$, after the cold mode has disappeared. At this point, the neutral fraction equals $1.4 \times 10^{-4}$.}
    \label{fig:phase-diagram} 
    \end{figure*}

Before the Epoch of Reionization, the IGM consists of mostly neutral hydrogen (as well as neutral helium) after the recombination of protons and electrons in the universe. But as gas clouds begin to condense and form stars and galaxies, these objects emit radiation that reionizes the IGM over time. The excess energy of ionizing photons is deposited as heat in the IGM, raising the temperature to a few tens of thousands of Kelvins, a typical temperature of the photoionized gas. 
This thermal evolution of the IGM is shown through the temperature-density phase diagram in Fig.~\ref{fig:phase-diagram}, represented at three different scale factors. 

The quantity on the $x$-axis of Fig.~\ref{fig:phase-diagram} is:
\begin{equation}
    1+\delta = 1+\frac{\rho-\overline{\rho}}{\overline{\rho}} = \frac{\rho}{\overline{\rho}},
    \label{eq:overdensity}
\end{equation}
where $\delta$ represents the overdensity, calculated as the local density $\rho$ minus the mean density of the universe $\overline{\rho}$, divided by the mean density of the universe, as shown by equation~(\ref{eq:overdensity}).

The diagram at three scale factors in Fig.~\ref{fig:phase-diagram}, which has an $x$-axis density bin width of $\log(1+\delta)=0.01$ and a $y$-axis temperature bin width of $\log(T/K)=0.01$, showcases how the gas is being heated from the ionizing radiation and is moving to a higher temperature, settling between $T = 10^{4}$ K and $T = 2\times 10^{4}$ K in the last panel. This warm gas is expected to be within the bubbles of ionized IGM, while the cold, neutral gas is expected to be outside of those ionized bubbles. 

While the underlying physical processes of the TDR are well understood, the resulting quantitative trends of the evolution are not. This is particularly because the Lyman-$\alpha$ forest observations used are difficult to obtain at higher redshifts corresponding to the Epoch of Reionization given the higher neutral fractions, and therefore these observations do not constrain the TDR during the Epoch of Reionization.
% The evolution of the temperature-density relation during reionization is not quantitatively well understood. 
We can investigate this evolution by defining the two gas ``modes'' (warm and cold) and finding relations between the timing of the transition from the cold to the warm mode and the timing of the overall reionization process.

\subsection{Mode Separation Methods}

\begin{figure*}
    \includegraphics[width=7in]{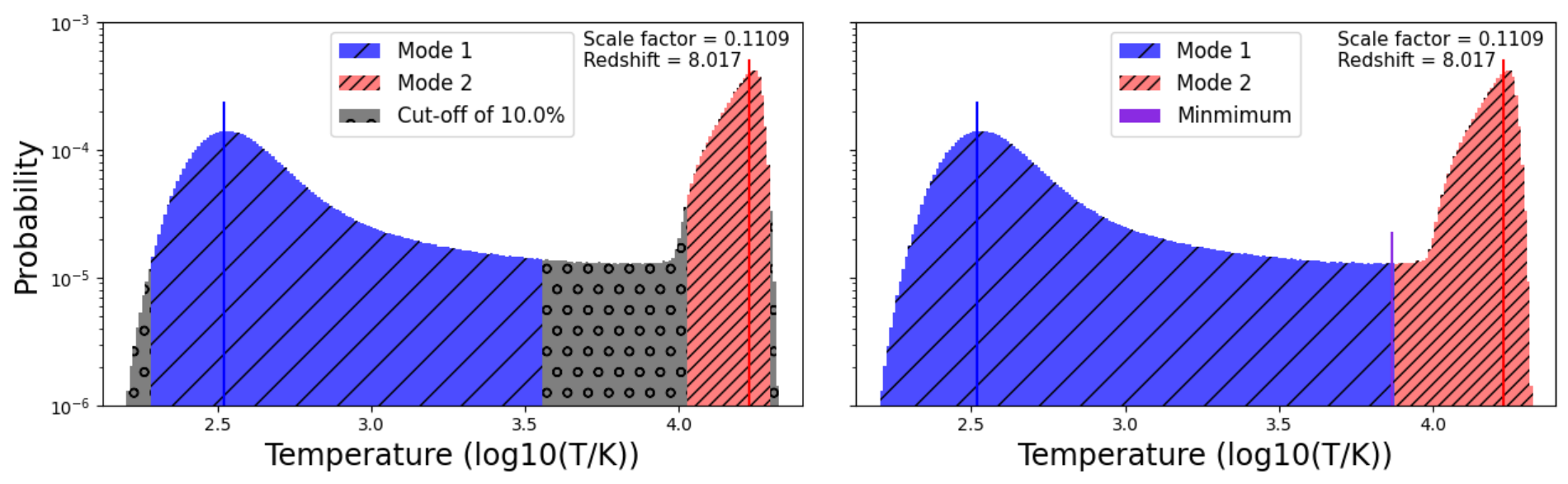}
    \caption{The separation of the two modes of the
    IGM as defined using two different methods, with the cut-off method (bins with values below 10\% of the peak value are not included in mode definitions) displayed in the first (left) panel and the minimum method (modes are separated by the bin with the lowest histogram value between them) displayed in the second (right) panel. The histogram is shown for gas at the mean cosmic density, $0<\log(1+\delta)<0.1$ for a simulation with intermediate reionization history 1.}
    \label{fig:mode-sep}
\end{figure*}

The two modes of the IGM can easily be seen in a one-dimensional histogram for gas temperature at a fixed density. An example of such a histogram is shown in Fig.~\ref{fig:mode-sep}. The leftmost peak of each panel of the figure represents the colder mode of neutral hydrogen, whereas the rightmost peak of each panel represents the warmer mode of reionized hydrogen. The modes are almost always distinct but are not completely separated. Because of this, we define two different methods for separating all or most of the cells in the simulation into the two modes.

The left panel of Fig.~\ref{fig:mode-sep} displays the mode separation by use of the cut-off method, where the modes are first separated by the $\sim$$10^{4} \, \mathrm{K}$ threshold for ionized hydrogen. For a peak above $\sim$$10^{4} \, \mathrm{K}$, we label the resulting mode the 'warm mode', and for a peak below $\sim$$10^{4} \, \mathrm{K}$, we label it the 'cold mode'. Then, we only count the bins that lie above a threshold (such as 10\% of the peak height, as shown in the left panel in Fig.~\ref{fig:mode-sep}). This is done to establish a more constrained mode definition that focuses on the peak values and a more limited range of their distributions. 
The right panel of Fig.~\ref{fig:mode-sep} displays the mode separation by use of the minimum method, where the modes are simply separated by the lowest histogram value between the two peaks. This method allows for a wider distribution in the mode definition, and it does not require the explicit $\sim$$10^{4} \, \mathrm{K}$ threshold. However, the temperature at which this minimum occurs is always near this threshold. 
Further analysis of these modes throughout the text suggests that the two mode separation methods yield indistinguishable results. See Appendix~\ref{app:mode_sep} to view the insignificant differences when comparing the two mode separation methods. 

\section{Results}

\subsection{Timing of the transition}

\begin{figure}
    \includegraphics[width=\columnwidth]{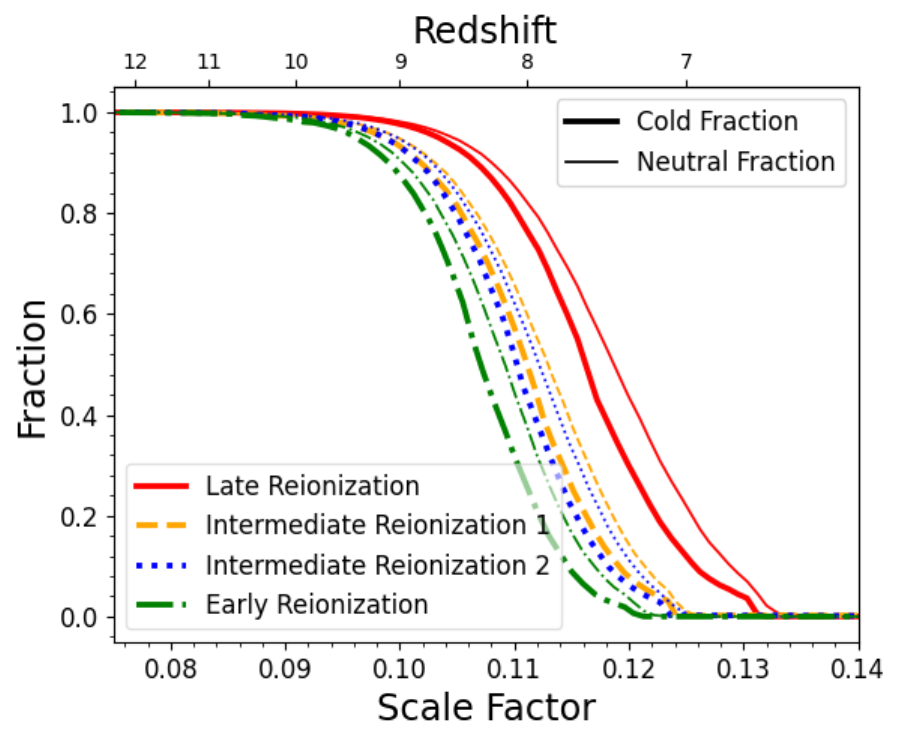}
    \caption{The fraction of the defined cold mode (thick lines) and the volume-weighted neutral fraction of the IGM (thin lines, the same as shown in Fig.~\ref{fig:rei-history}) as a function of scale factor, using the minimum method (modes are separated by the minimum histogram value between the two mode peaks), shown in simulation boxes with different reionization histories.}
    \label{fig:rei-hist-mode}
\end{figure}

By assigning all or most of the cells in the simulation to one of the modes, we can calculate the cold mode fraction with respect to the sum of both the warm and cold modes. Fig.~\ref{fig:rei-hist-mode} displays this cold mode fraction as well as the volume-weighted neutral fraction (the same as in Fig.~\ref{fig:rei-history}) over time for each box. The cold mode fraction has a similar evolution as the neutral fraction, but it transitions slightly earlier than the neutral fraction. This is because higher energy photons of the ionizing radiation have a longer mean free path (MFP) and will heat the neutral gas before the more numerous lower energy ionizing photons can fully ionize it. Because the reionization process is dominated by these low-energy photons and heating is more sensitive to higher-energy photons, any particular place in the IGM is heated somewhat before being fully ionized.                                                                                             

\begin{figure}
    \includegraphics[width=\columnwidth]{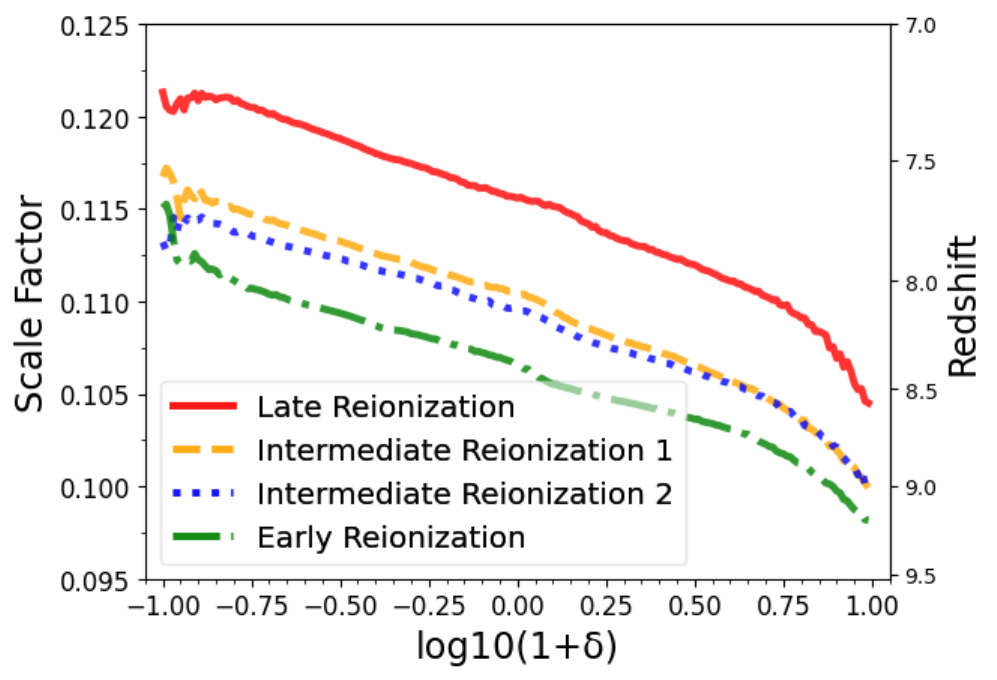}
    \caption{The scale factors, for each density bin, at which the lower temperature (cold mode) and higher temperature (warm mode) fractions are approximately equal, and thus there are approximately equal amounts of gas in either mode, using the minimum method (modes are separated by the minimum histogram value between the two mode peaks), shown in simulation boxes with different reionization histories.}
    \label{fig:critical-dense}
\end{figure}

From Fig.~\ref{fig:rei-hist-mode} we can see that the cold mode fraction equals 0.50 at some scale factor, which is the point where the cold and warm mode fractions are equal. This signifies the moment when the warm mode begins overtaking the cold mode in volume. We call the scale factor at which this occurs the ``critical scale factor'', and this value can be used as a quantitative measure of the timing of the cold-to-warm mode transition in each density bin. Fig.~\ref{fig:critical-dense} displays the critical scale factor across the density bins for each simulation box. 

\begin{figure}

	\includegraphics[width=\columnwidth]{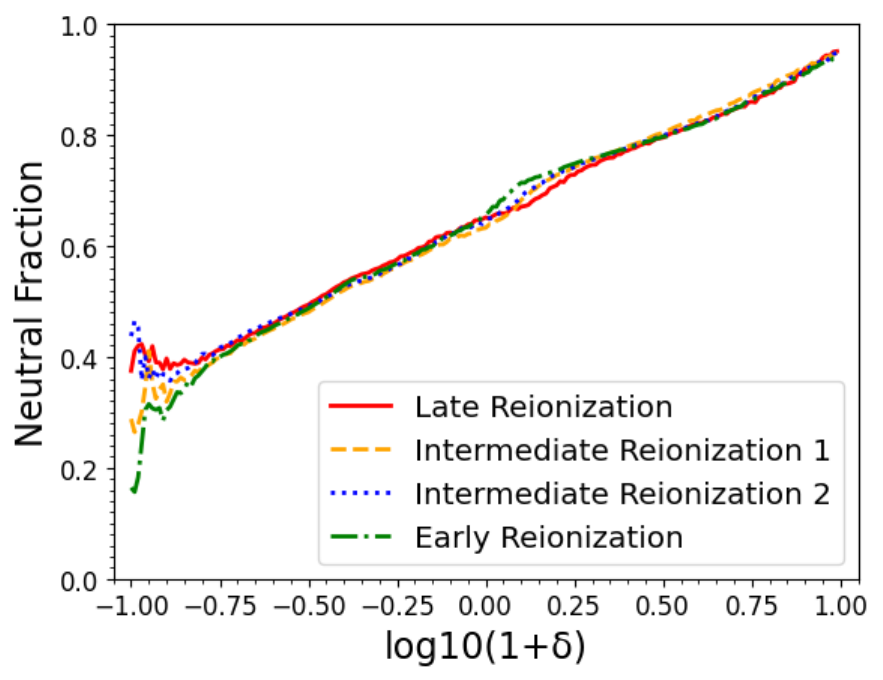}
    \caption{The neutral fraction of the IGM for each density bin at the time when the defined cold and warm mode fractions are equal, using the minimum method (modes are separated by the minimum value between the two mode peaks). At this point, the amount of gas in the `warm' mode begins surpassing the 'cold' mode.} 
    \label{fig:neutral-critical}

\end{figure}

The heating and the ionization of the IGM are two aspects of the same process, and hence one may wonder whether the connection between them is ``universal'', i.e.\ independent of the details of the actual reionization history. In order to explore that, we show in Fig.~\ref{fig:neutral-critical} a variation of Fig.~\ref{fig:critical-dense} with values for critical scale factors replaced by the values of the volume-weighted neutral fraction at that moment for that simulation box. 

Different simulation boxes (with their differing ionization histories) at each density have near-universal neutral fractions at the critical scale factors for the cold modes. In other words, when the thermal history is parameterized by the value of volume-weighted neutral fraction, the transition from cold to warm mode becomes almost universal. Note that this conclusion also holds if we separate the cold and warm modes via the cut-off method (see Appendix~\ref{app:mode_sep}). We elaborate on this universality in section~\ref{sec:conclusions}.

\subsection{Emergence of a temperature-density relation}
\label{sec:temp-dense}

The tight temperature-density relation is perhaps the most recognizable feature of the Lyman-$\alpha$ forest at intermediate redshifts. Fig.\ \ref{fig:phase-diagram} shows that such a relation is not present during the Epoch of Reionization, and hence it must ``emerge'' shortly after reionization is completed.  In order to determine the emerging temperature-density relation $T(\delta) = T_0(1+\delta)^{\gamma-1}$, we extract the most likely temperature at each density in the warm mode - i.e.\ the peak of the temperature histogram in each density bin, shown with vertical lines within the rightmost (warm) modes in Fig.~\ref{fig:mode-sep}. 

\begin{figure*}
    \includegraphics[width=7in]{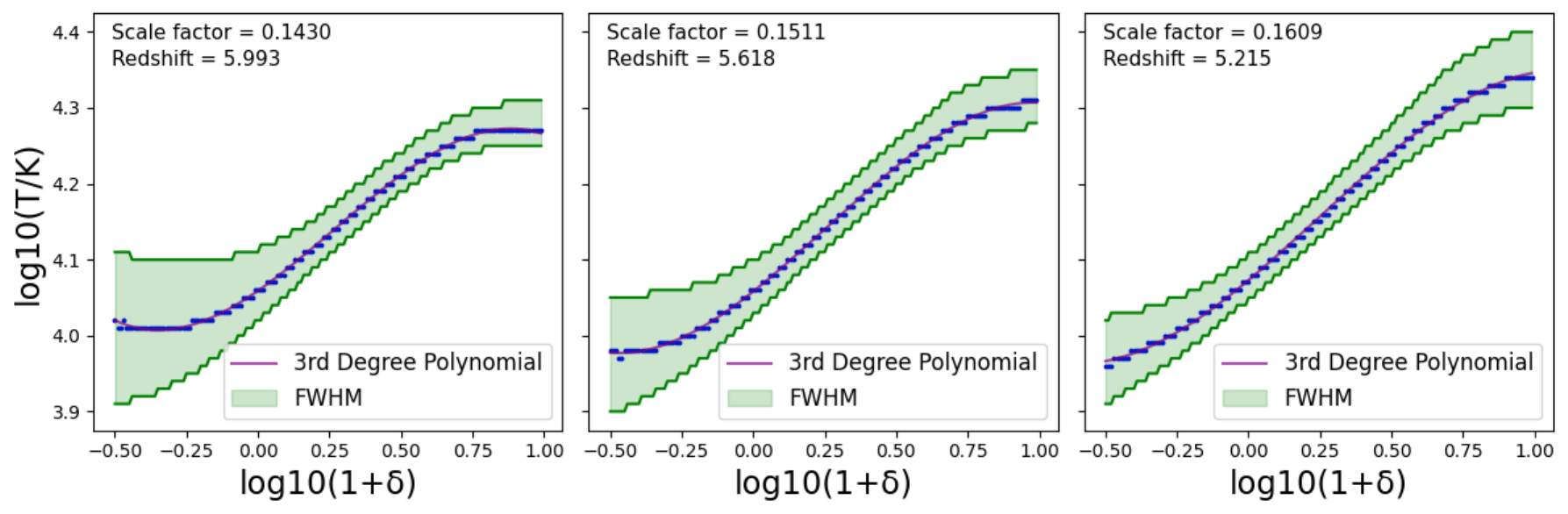}
    \caption{The most likely temperature (the peak of the temperature histogram) for each density at three different scale factors (points), with the full width at half maximum (FWHM) of the warm mode also shown (as a surrounding band) for the intermediate reionization 1 box. This shows the decreasing evolution of the FWHM once the temperature-density relation has been established. The power-law slope at the mean density is measured using a 3rd-degree polynomial on a log-log scale for the whole density range shown (lines).} 
    \label{fig:peakTemp}
\end{figure*}

The parameters of the temperature-density relation, the temperature at the mean density $T_0$ and the power-law slope $\gamma$, can be measured by fitting a power-law to a range of densities around the mean cosmic density. This procedure is not well defined, though, as it depends on the specific density range chosen. Instead, we noticed that the full temperature-density relation is well fit by a 3rd-degree polynomial in log-log space, which we can then use to determine $T_0$ and the slope $\gamma$ (at the mean density) accurately. The ``emergence'' of the temperature-density relation is not quantified by its parameters, however, but rather by how tight it is. In order to quantify that, we also measure the full-width-at-half-maximum (FWHM) of the warm mode around the peak temperature. Fig.~\ref{fig:peakTemp} shows the most likely temperature of the warm mode as well as its FWHM at three different scale factors.

\begin{figure*}
	\includegraphics[width=7in]{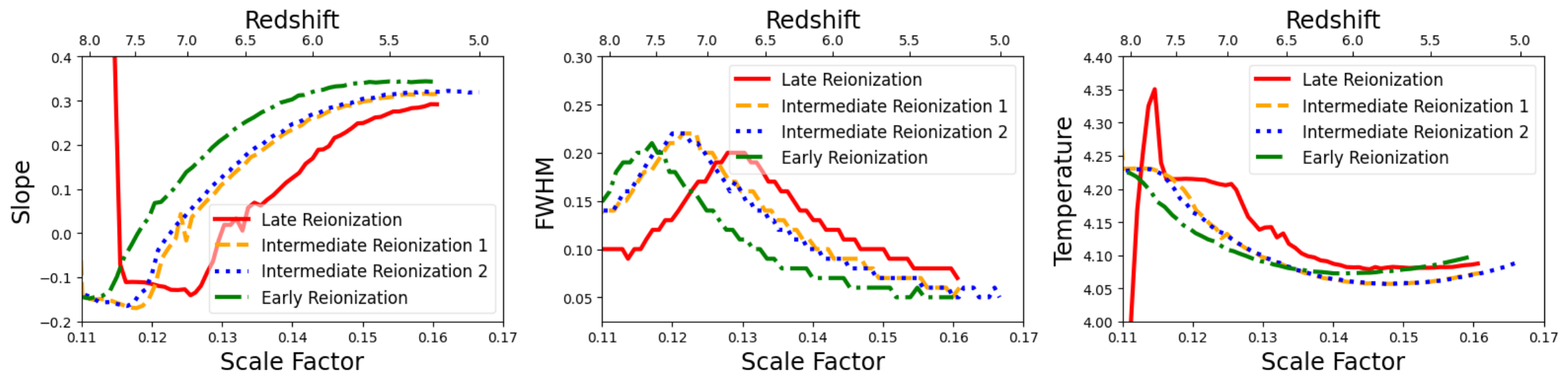}
    \caption{The slope, the full-width-at-half-maximum (FWHM) of the temperature distribution at the mean density ($0<\log(1+\delta)<0.1$ bin), and the most likely temperature at the mean density $T_0$ as a function of scale factor (as well as redshift). The redshift axis is shown on the top $x$-axis.} 
    \label{fig:slope-FWHM-a}
\end{figure*}

\begin{figure*}
	\includegraphics[width=7in]{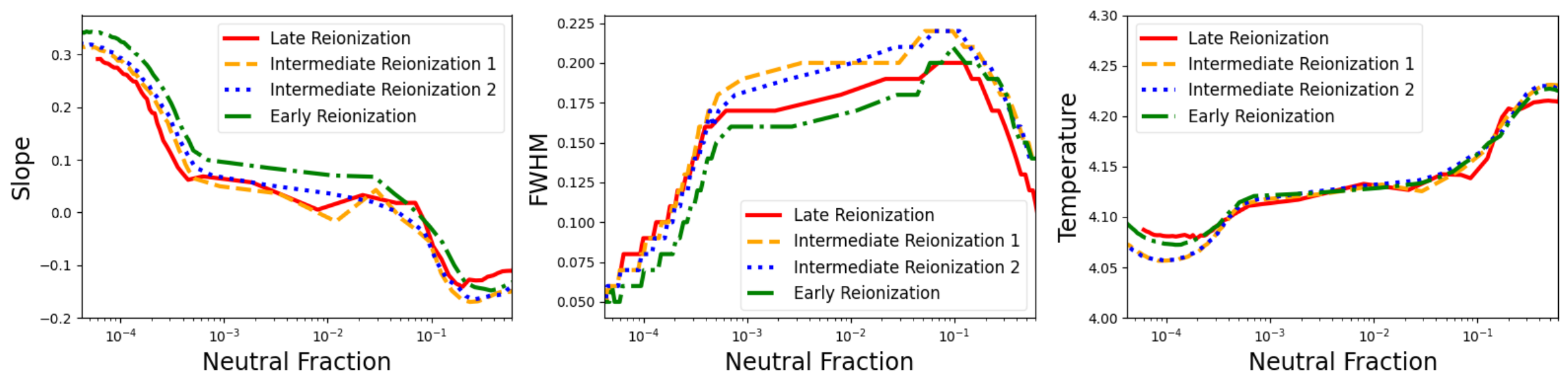}
    \caption{The same as Fig.\ \protect{\ref{fig:slope-FWHM-a}}, now as a function of the volume-weighted neutral fraction. The emergence of the tight temperature-density relation (i.e.\ a rapid decrease in the width of the distribution) occurs when the neutral fraction is $10^{-4}\lesssim X_{HI}\lesssim10^{-3}$. %0<log10(1+d)<0.1. 
    }
    \label{fig:slope-FWHM-x}
\end{figure*}

From Fig.~\ref{fig:peakTemp}, we can see that the FWHM of the warm mode in the mean density bin $0<\log(1+\delta) < 0.01$ is decreasing as the scale factor increases, illustrating the emergence of the temperature-density relation over time. We see for each scale factor shown that the FWHM is greater at lower densities than near the mean density, confirming that as the low-density gas is transitioning from cold to warm later than gas near the mean density (see Fig.~\ref{fig:critical-dense}), the temperature-density relation also emerges later in the low-density gas. 

The behavior of the slope and the FWHM at the mean density (which we use to characterize the ``emergence'' of the TDR) can also be investigated over time or with respect to the neutral fraction, shown in Fig.~\ref{fig:slope-FWHM-a} and Fig.~\ref{fig:slope-FWHM-x}, respectively. From the differences in the simulation boxes, we can see that in Fig.~\ref{fig:slope-FWHM-a}, earlier reionization leads to a consistently steeper slope and smaller FWHM than later reionization does. From the evolution of the slope and FWHM across neutral fractions shown in Fig.~\ref{fig:slope-FWHM-x}, we again observe an approximate universality among the different simulation boxes, i.e., among the different reionization histories. From this figure, we can see that this universality is not exact and is not as precise as the universality for the timing of the cold-to-warm mode transition (Fig.~\ref{fig:neutral-critical}). Additionally, the uncertainty in these lines has not been quantified, so we cannot say whether the differences between the reionization histories are larger than the inherent uncertainty for the individual histories (due to the inhomogeneous nature within a single simulation box). Multiple factors can affect this slight non-universality, from actual physical non-universality, to numerical artifacts in the simulations, to particular numerical choices in our analysis (such as binning). At this point it does not seem worthwhile to explore these factors - the fact that the (near) universality exists is much more important than the reasons for it not being perfect.  In particular, Fig.~\ref{fig:slope-FWHM-x} shows that the emergence of the temperature-density relation primarily occurs when the neutral fraction is $10^{-4}\lesssim X_{HI}\lesssim10^{-3}$ irrespective of other details of reionization.

\section{Discussion and Conclusions}
\label{sec:conclusions}

In this paper, we investigate the evolution of the thermal state of the IGM during cosmic reionization. Our primary goal is to better understand how the reionization history impacts the emergence and the evolution of the tight temperature-density relation, so ubiquitous in the lower redshift Lyman-$\alpha$ forest.

Our main finding is that the emergence of the temperature-density relation is nearly ``universal'', in the sense that when expressed as a function of the volume-weighted neutral fraction instead of the actual time itself, both timing and the shape of the temperature-density relation become only weakly dependent on the details of ionization history. In particular, the timing of the transition from the cold neutral IGM into the warm ionized IGM (Fig.\ \ref{fig:neutral-critical}) is almost perfectly universal at all densities. This fact is not necessarily surprising, since ionized gas is warm, but it is not trivial either. First, not all warm gas is ionized as Fig.~\ref{fig:rei-hist-mode} shows (the heating history precedes the ionization history). Second, and most importantly, Fig.~\ref{fig:neutral-critical} relates the cold-to-warm transition at a given density to the global volume-weighted neutral fraction, and hence depends on the overall density distribution. Hence, the (near) universality \emph{at a fixed density} is not necessarily assured. The mean neutral fraction depends on the full ranges of densities. Hence, the cold-to-warm transition at a fixed density does not have to be correlated with the mean neutral fraction. For example, if the density PDF evolves faster at higher densities than at lower densities, the mean neutral fraction would evolve faster than the cold-to-warm transition at lower densities. The fact that this is not the case and the universality is highly accurate, implies that the timing of reionization is in some way ``synchronized'' with the evolution of the density PDF - an idea that has not been clearly stated or explored previously. More than that, we observe the universality between the 4 distinct $40h^{-1}$ Mpc boxes, implying that the ``synchronization'' between reionization and the evolution of the density PDF occurs on scales at least as small as several tens of cMpc.

In order to gain some intuition on how such a ``synchronization'' may develop, let us consider a toy model of a non-evolving density field. Let us also imagine, merely for the sake of the argument, that reionization proceeds totally randomly. Then, half of the gas at a given density would become warm when half of the universe is ionized, i.e.\ the timing would be universal, but the dependence on the mean neutral fraction would be trivial, all lines in Fig.\ \ref{fig:neutral-critical} would be strictly horizontal. Notice that accounting for the finite width of ionization fronts would not solve the difference, but rather exacerbate it: since the mean free path is longer in lower-density gas, the lower-density gas would get heated at slightly higher values of the mean neutral fractions than the denser gas. i.e., in the random ionization scenario lines in Fig.\ \ref{fig:neutral-critical} would slope slightly downward.

A much more realistic model for reionization is a barrier-crossing formalism of \citet{Furlanetto2004}. In that model, the lower-density gas gets ionized later than the higher-density gas. Hence, there is indeed a positive correlation between the neutral fraction at the time of the cold-to-warm transition at fixed density and the value of density. However, since different $40h^{-1}$ Mpc have different density PDFs, the dependence shown in Fig.\ \ref{fig:neutral-critical} would not be universal among our 4 simulation boxes. In this paper, we do not offer an alternative scenario that explains Fig.\ \ref{fig:neutral-critical} - understanding the universality will require a significantly larger effort, and a focus significantly different from the subject of this paper, and hence we leave such effort for future work. 

The near universality in the evolution of the temperature-density relation of the warm mode (the only mode where the well-defined temperature-density relation actually exists) is less striking than the near universality of the timing of the cold-to-warm transition, but it is still significant (Fig.~\ref{fig:slope-FWHM-x}). This allows us to make generic conclusions about the emergence of the tight temperature-density relation in reionization-history-independent terms. For example, the emergence primarily occurs when the volume-weighted neutral fraction decreases from $10^{-3}$ to $10^{-4}$.

The (near) universality is a prediction that is easily testable by observations: since different regions of the universe reionize at different times, observations of absorption spectra of different quasars should show this near universality - or demonstrate a major shortcoming of the CROC project simulations.

\section*{Acknowledgements}

AW would like to thank the National Science Foundation for funding the University of Michigan Research Experiences for Undergraduates program (Grant No. \#2149884). This work was supported in part by the NASA Theoretical and Computational Astrophysics Network (TCAN) grant 80NSSC21K0271.
This manuscript has been co-authored by Fermi Research Alliance, LLC under Contract No. DE-AC02-07CH11359 with the U.S. Department of Energy, Office of Science, Office of High Energy Physics. This work used resources of the Argonne Leadership Computing Facility, which is a DOE Office of Science User Facility supported under Contract DE-AC02-06CH11357. An award of computer time was provided by the Innovative and Novel Computational Impact on Theory and Experiment (INCITE) program. This research is also part of the Blue Waters sustained-petascale computing project, which is supported by the National Science Foundation (awards OCI-0725070 and ACI-1238993) and the state of Illinois. Blue Waters is a joint effort of the University of Illinois at Urbana-Champaign and its National Center for Supercomputing Applications. This work was completed in part with resources provided by the University of Chicago’s Research Computing Center.
DR acknowledges funding provided by the Leinweber Graduate Fellowship at the University of Michigan.
We would lastly like to thank the anonymous referee for constructive and helpful comments.
%%%%%%%%%%%%%%%%%%%%%%%%%%%%%%%%%%%%%%%%%%%%%%%%%%
\section*{Data Availability}

% The inclusion of a Data Availability Statement is a requirement for articles published in MNRAS. Data Availability Statements provide a standardised format for readers to understand the availability of data underlying the research results described in the article. The statement may refer to original data generated in the course of the study or to third-party data analysed in the article. The statement should describe and provide means of access, where possible, by linking to the data or providing the required accession numbers for the relevant databases or DOIs.
The phase diagram and reionization history data for the four simulation boxes used in this paper can be made available upon request.

The code used to display all plots in this work can be found within the GitHub repository: \url{https://github.com/wellsalexandra/IGM-Phase}.



\bibliographystyle{mnras}
\bibliography{main}

%%%%%%%%%%%%%%%%%%%%%%%%%%%%%%%%%%%%%%%%%%%%%%%%%%

%%%%%%%%%%%%%%%%% APPENDICES %%%%%%%%%%%%%%%%%%%%%

\appendix

\section{Indistinguishable Results from Mode Separation Methods}
\label{app:mode_sep}

\begin{figure}
	\includegraphics[width=\columnwidth]{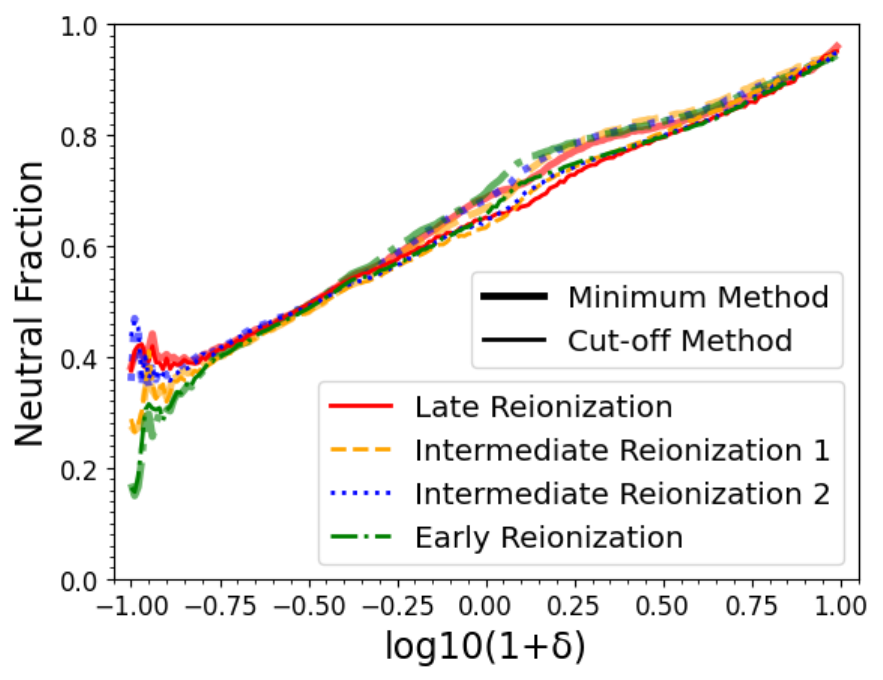}
    \caption{The neutral fraction of the IGM for each density bin at the time when the mode fractions are approximately equal. The thin line represents the cut-off mode separation method (the bottom 10\% of each mode is cut off), and the thick line represents the minimum method (modes are separated by the minimum histogram value between the two mode peaks). The two methods do not show any notable differences in their resulting analyses.}
    \label{fig:neutral-dense-both}
\end{figure}

Fig.~\ref{fig:neutral-dense-both} is a version of Fig.~\ref{fig:neutral-critical} that shows both mode separation methods. Because the two methods do not yield notable differences in their results, we resolved to only show this plot for one method in Fig.~\ref{fig:neutral-critical}. 

%%%%%%%%%%%%%%%%%%%%%%%%%%%%%%%%%%%%%%%%%%%%%%%%%%

% Don't change these lines
\bsp	% typesetting comment
\label{lastpage}
\end{document}